\documentclass[floatfix,aps,pra,twocolumn,showpacs,superscriptaddress,10pt]{revtex4-2}
\usepackage[utf8]{inputenc}
\usepackage[T1]{fontenc}
\usepackage{color}
\usepackage{float}
\usepackage{mathrsfs}
\usepackage{amstext}
\usepackage{graphicx}
\usepackage{amsmath}
\usepackage{amstext}

\makeatletter
\usepackage{lipsum}

\makeatother

\usepackage{babel}
\begin{document}
\title{Steady-state entanglement generation for non-degenerate qubits}

\author{Murilo H. Oliveira}
\email{murilo.oliveira@df.ufscar.br}
\affiliation{Departamento de Física, Universidade Federal de São Carlos, P.O. Box
676, 13565-905, São Carlos, São Paulo, Brazil.}

\author{Gerard Higgins}
\thanks{Present address: Institute for Quantum Optics and Quantum Information, Austrian Academy of Sciences, Vienna, Austria}
\affiliation{Department of Physics, Stockholm University, 10691 Stockholm, Sweden}

\author{Chi Zhang}
\thanks{Present address: Division of Physics, Mathematics, and Astronomy, California Institute of Technology,
Pasadena, CA 91125, USA}
\affiliation{Department of Physics, Stockholm University, 10691 Stockholm, Sweden}

\author{Ana Predojević}
\affiliation{Department of Physics, Stockholm University, 10691 Stockholm, Sweden}

\author{Markus Hennrich}
\affiliation{Department of Physics, Stockholm University, 10691 Stockholm, Sweden}

\author{Romain Bachelard}
\affiliation{Departamento de Física, Universidade Federal de São Carlos, P.O. Box
676, 13565-905, São Carlos, São Paulo, Brazil.}
\affiliation{Universit\'e C\^ote d'Azur, CNRS, Institut de Physique de Nice, 06560 Valbonne, France}

\author{Celso J. Villas-Boas}
\affiliation{Departamento de Física, Universidade Federal de São Carlos, P.O. Box
676, 13565-905, São Carlos, São Paulo, Brazil.}

\begin{abstract}
We propose a scheme to dissipatively produce steady-state entanglement in a two-qubit system, via an interaction with a bosonic mode. The system is driven into a stationary entangled state, while we compensate the mode dissipation by injecting energy via a coherent pump field. We also present a scheme which allows us to adiabatically transfer all the population to the desired entangled state. The dynamics leading to the entangled state in these schemes can be understood in analogy with electromagnetically induced transparency (EIT) and stimulated Raman adiabatic passage (STIRAP), respectively.   

\end{abstract}

\maketitle

\section{Introduction}

Entanglement is the clearest nonclassical signature of quantum physics. A composite system is considered to be entangled when the quantum state that describes it is inseparable, \textit{i.e.},  it is impossible to write it as a product of the states of each subsystem \cite{horodecki2009}. In the last decades, entangled states have been the subject of great interest, presenting themselves as a resource for several quantum schemes and applications such as quantum communication \cite{ekert1991,bennett1993,bennett2000}, quantum computation \cite{gottesman1999}, metrology \cite{giovannetti2006} and quantum sensing \cite{degen2017}.

The success of the aforementioned applications and tests often depends on the ability to generate long-lived entangled states. However, in a realistic situation, the system will interact with the environment. This will inevitably lead to the deterioration of the entangled state, which is sensitive to decoherence \cite{horodecki2009}. For this reason, entanglement preservation schemes have gained great prominence. Among the proposed methods to minimize unwanted decoherence, we should mention the use of decoherence-free subspaces \cite{lidar1998,viola1998}, quantum error correction codes \cite{steane1996,knill1997}, weak measurements \cite{sun2010,kim2012} and the quantum Zeno effect \cite{facchi2004,maniscalco2008}. 

Instead of aiming to prevent decoherence, a different strategy involves engineering the system-environment interaction to \textit{generate} entangled states, these are called dissipation-assisted entanglement generation methods \cite{diehl2008,kraus2008,verstraete2009}. Since this idea was presented, numerous implementations have been proposed and experimentally realized using several physical platforms, such as cavity QED \cite{plenio1999,nicolosi2004,wang2010,kastoryano2011,del2011}, superconducting qubits \cite{shankar2013,kimchi2016,liu2016,doucet2020}, macroscopic atomic ensembles \cite{krauter2011,santos2022}, Rydberg atoms \cite{rao2013,carr2013,shao2017} and trapped ions \cite{barreiro2011,lin2013,bentley2014,horn2018,cole2021,cole2022}. Another widely-studied approach is the creation of long-lived entangled states via stimulated Raman adiabatic passage (STIRAP) \cite{vitanov2017,bergmann2019}, since it offers robustness against decoherence by not populating lossy states.

In this paper we propose two new schemes for producing highly-entangled states in a system of two non-degenerate qubits. It is known that, in some systems, it's possible to manipulate the degree of entanglement between two qubits via Stark shifts of their electronic levels \cite{hettich2002,wang2010,trebbia2021}. Here, however, we show that the symmetry in the energy shifts between the emitters with respect to the bosonic mode actually allows us to achieve a stronger entanglement. By considering their effective interaction through a bosonic mode, such as an optical cavity or a motional mode, we are able to achieve a stronger coupling between the qubits without needing to place the qubits particularly close together. This interaction with the quantized mode provides a coupling regime strong enough so that the time scales of the effective interactions are much faster than the qubit relaxation, leading to higher degrees of entanglement. This shows that the distinguishability between the quantum emitters can be an advantage in the quest for the producing highly-entangled states. We are able to achieve a maximally-entangled steady state, which is maintained by injecting power via a pump field. 

We show that the dynamics leading to the highly-entangled two-qubit state can be understood by comparison with electromagnetically induced transparency (EIT) regime \cite{fleischhauer2005}. In that same analogy, but restricting ourselves to the subspace of just a single excitation, we are able to drive the system into an entangled state via a STIRAP-like process. 

\section{Model}

\begin{figure}[ht]
\begin{centering}
\includegraphics[width=0.9\columnwidth]{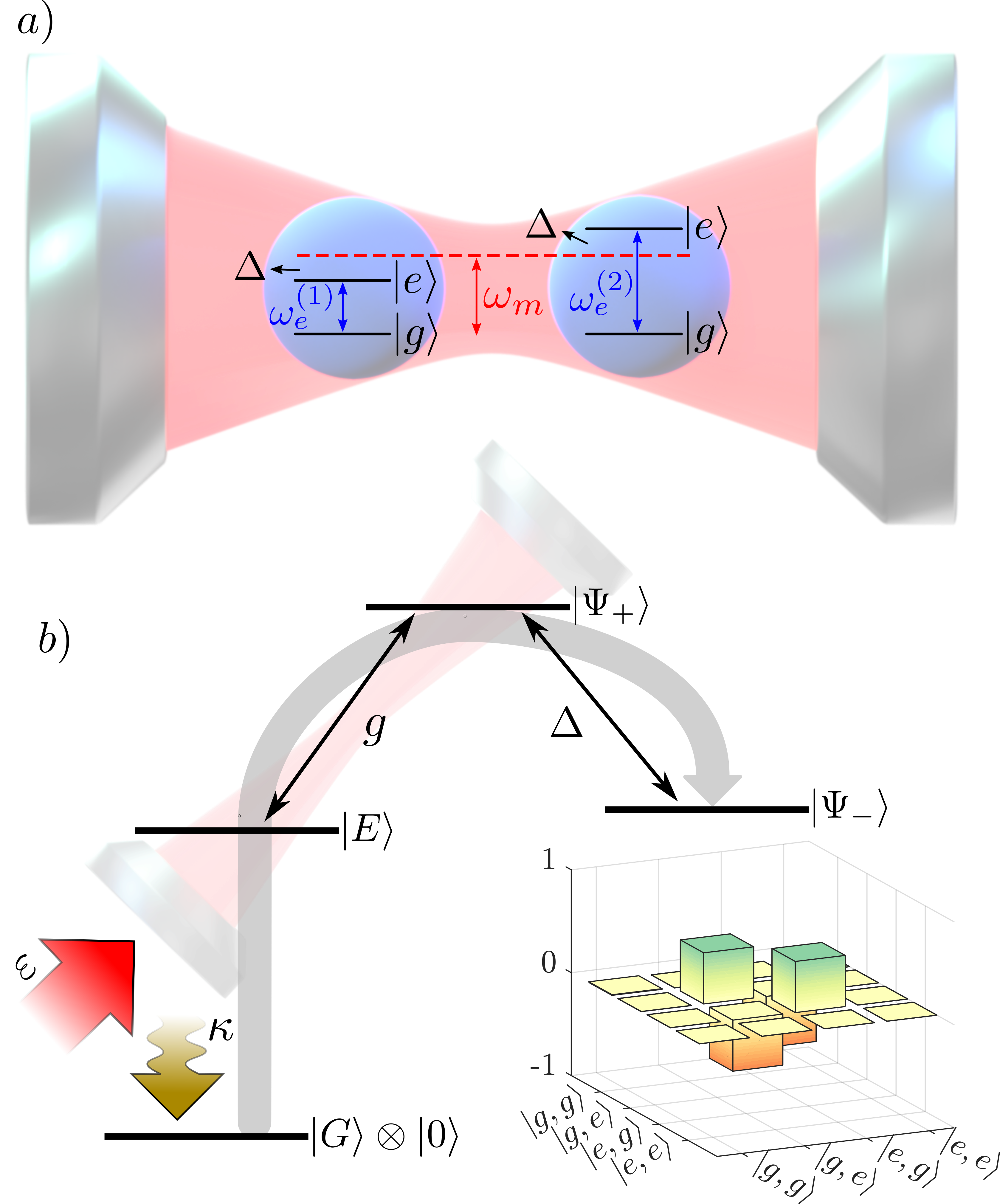}
\par\end{centering}
\caption{a)~Two non-degenerate qubits, with both ground states coupled to the bosonic mode of frequency $\omega_{m}$, detuned by $\pm \Delta$. b) Level scheme of the same system, but now in the basis up to one excitation:  $\vert G\rangle \otimes \vert 0\rangle$, $\vert E\rangle=\vert G\rangle \otimes \vert 1\rangle$, and $\vert\Psi_{\pm}\rangle=\vert\Phi_{\pm}\rangle\otimes\vert 0 \rangle$, where $\vert G\rangle=\vert g,g \rangle$ and $\vert \Phi_{\pm}\rangle = \left(\vert e,g\rangle\pm\vert g,e\rangle\right)/\sqrt{2}$. Here, $g$ promotes transitions from $\vert E\rangle$ to $\vert\Psi_{+}\rangle$, while $\Delta$ from $\vert\Psi_{+}\rangle$ to $\vert\Psi_{-}\rangle$. We consider a pump field, of strength $\varepsilon$, continuously injecting energy into the mode to combat decay from the mode with rate $\kappa$. The inset in b) shows the entangled steady state partial density matrix, where the bosonic mode has been traced out.}
\label{fig:1}
\end{figure}

Let us consider a system of two qubits, with different resonance frequencies $\omega_e^{(1)}$ and $\omega_e^{(2)}$, which are coupled to the same bosonic mode with frequency $\omega_m$, as illustrated in Fig.~\ref{fig:1}(a). Here, the bosonic mode is symmetrically detuned from each of the  qubits, so that $\omega_e^{(1)}=\omega_m-\Delta$ and $\omega_e^{(2)}=\omega_m+\Delta$. In the Schr\"odinger picture, the system Hamiltonian reads ($\hbar=1$):
\begin{eqnarray}
\hat{H} = &\omega_{e}^{(1)}\hat{\sigma}_{ee}^{(1)}+\omega_{e}^{(2)}\hat{\sigma}_{ee}^{(2)}+\omega_{m}\hat{a}^{\dagger}\hat{a}\nonumber\\
&+g\left[\hat{a}\left(\hat{\sigma}_{+}^{(1)}+\hat{\sigma}_{+}^{(2)}\right)+h.c.\right],\label{eq:1}
\end{eqnarray}
where $\hat{\sigma}_{+}^{(k)}=\vert e \rangle\langle g\vert$, $\hat{\sigma}_{-}^{(k)}=\vert g\rangle\langle e\vert$ and $\hat{\sigma}_{ee}^{(k)}=\vert e\rangle\langle e\vert$ are the raising, lowering and excited-state population operators, respectively, acting on the $k$-th qubit (with $k\in\{1,2\}$). Without loss of generality, we define the ground state energy to be zero. $\hat{a}$ ($\hat{a}^\dagger$) is the annihilation (creation) operator of the bosonic mode, $g$ is the coupling strength between the bosonic mode and each of the qubits, and $h.c.$ stands for the Hermitian conjugate. 

For convenience, we move to the interaction picture, make the rotating wave approximation and move to a rotating referential of relative coordinates in which both qubits are stationary, thus eliminating the Hamiltonian time dependence. Then, Eq. \eqref{eq:1} becomes
\begin{equation}
\mathcal{\hat{H}}=\Delta\left(\hat{\sigma}_{ee}^{(1)}-\hat{\sigma}_{ee}^{(2)}\right)+g\left[\hat{a}\left(\hat{\sigma}_{+}^{(1)}+\sigma_{+}^{(2)}\right)+h.c.\right].\label{eq:3}
\end{equation}

To account for decoherence, we consider our system to be a weak system-environment coupling regime, which allows us to use the Lindblad master equation \cite{petruccione2002} at temperature $T=0\,$K. The assumption of zero temperature is reasonable since we work within the optical regime, where the number of thermal photons remain negligible even for room temperatures. Thus, we obtain the dynamical equations for the density matrix $\rho$ 
\begin{equation}
\dot{\hat{\rho}}=-i\left[\hat{\mathcal{H}},\hat{\rho}\right]+\hat{\mathscr{L}}_{\mathrm{q}}^{(1)}+\hat{\mathscr{L}}_{\mathrm{q}}^{(2)}+\hat{\mathscr{L}}_{m},\label{eq:5}
\end{equation}
where 
\begin{equation}
\hat{\mathscr{L}}_{\mathrm{q}}^{(k)}=\Gamma\left(2\hat{\sigma}_{-}^{(k)}\hat{\rho}\hat{\sigma}_{+}^{(k)}-\hat{\sigma}_{ee}^{(k)}\hat{\rho}-\hat{\rho}\hat{\sigma}_{ee}^{(k)}\right)
\end{equation}
is the Lindblad term that accounts for the spontaneous decay from the excited state of the $k$-th qubit, with $k\in\{1,2\}$ and $\Gamma$ is the qubit decay rate, here assumed the same for both qubits. The Lindblad term
\begin{equation}
\hat{\mathscr{L}}_{m}=\kappa\left(2\hat{a}\hat{\rho} \hat{a}^{\dagger}-\hat{a}^{\dagger}\hat{a}\hat{\rho}-\hat{\rho} \hat{a}^{\dagger}\hat{a}\right)
\end{equation}
accounts for decay of the bosonic mode, where $\kappa$ is the decay rate.

\section{Steady state entanglement production}\label{sec2}
To describe the main mechanism responsible for the generation of entanglement, we restrict ourselves, for the moment, to the single-excitation subspace, which is composed of the following three states: $\vert E\rangle=\vert G\rangle \otimes \vert 1\rangle$ and $\vert\Psi_{\pm}\rangle=\vert\Phi_{\pm}\rangle\otimes\vert 0 \rangle$, where $\vert G\rangle=\vert g,g \rangle$ and $\vert \Phi_{\pm}\rangle = \left(\vert e,g\rangle\pm\vert g,e\rangle\right)/\sqrt{2}$. $\vert\Phi_{\pm}\rangle$ are maximally-entangled two-qubit states. In this subspace, the reduced Hamiltonian is: 
\begin{equation}
    \hat{\bar{\mathcal{H}}}=\Delta\vert\Psi_{-}\rangle\langle\Psi_{+}\vert +\sqrt{2}g\vert\Psi_{+}\rangle\langle E\vert +h.c.
    \label{eq:4}
\end{equation}
In analogy with the typical three-level $\Lambda$ systems, we consider that $\vert E\rangle$ and $\vert \Psi_-\rangle$ play the roles of the two ground states, while $\vert \Psi_+\rangle$ is the excited state, as depicted in Fig.~\ref{fig:1}(b). According to Eq.~\eqref{eq:4}, the transitions $\vert E\rangle \leftrightarrow \vert \Psi_+\rangle$ and $\vert \Psi_-\rangle \leftrightarrow \vert \Psi_+\rangle$ have effective coupling strengths $\sqrt{2}g$ and $\Delta$, respectively. The Hamiltonian in Eq.~\eqref{eq:4} has a \textit{dark eigenstate} (an eigenstate without a $\vert \Psi_{+}\rangle$ component) given by 
\begin{equation}
    \vert D\rangle = -\frac{\Delta}{\sqrt{\Delta^2+2g^2}}\vert E\rangle+\frac{\sqrt{2}g}{\sqrt{\Delta^2+2g^2}}\vert\Psi_{-}\rangle. \label{eq:darkstate}
\end{equation}
Just as in other three-level systems \cite{fleischhauer2005}, when the condition $g\gg\Delta$ is fulfilled, the dark state $\vert D\rangle$ transforms to $\vert\Psi_{-}\rangle$, which is our maximally-entangled target state. 

A significant difference between our system and a typical three-level $\Lambda$-system is the fact that neither $\vert E \rangle$ or $\vert\Psi_-\rangle$ are actually ground states. Because of this, the system keeps spontaneously decaying to the true zero energy ground state $\vert G\rangle \otimes \vert 0\rangle$. For this reason, the analogy becomes more accurate as the mode dissipation rate $\kappa$ and the spontaneous decay rate $\Gamma$ become negligible ($\kappa,\Gamma \ll g,\Delta$). Then the system effectively remains in the one-excitation subspace. One way to circumvent the decay and maintain the system in the one-excitation subspace is to keep injecting energy into the mode. We consider this energy injection as an additional term 
\begin{equation}
    \hat{H}_{\mathrm{pump}}=\varepsilon\left(\hat{a}^{\dagger}+\hat{a}\right) 
\end{equation}
of the system's Hamiltonian, where $\varepsilon$ is the the pump strength. In our simulations, $\varepsilon$ was kept constant at $\varepsilon=\kappa$.

To characterize the steady-state entanglement between the qubits, we choose the monotone quantifier concurrence \cite{Wootters1997,Wootters1998}. We numerically simulate the full system dynamics, taking into account the decoherence and also higher excited states. The concurrence is derived from the steady-state density matrix, which we obtain using the Quantum Toolbox in Python (QuTiP) \cite{qutip2013}, after the mode is traced out.
\begin{figure}[ht]
\begin{centering}
\includegraphics[width=1\columnwidth]{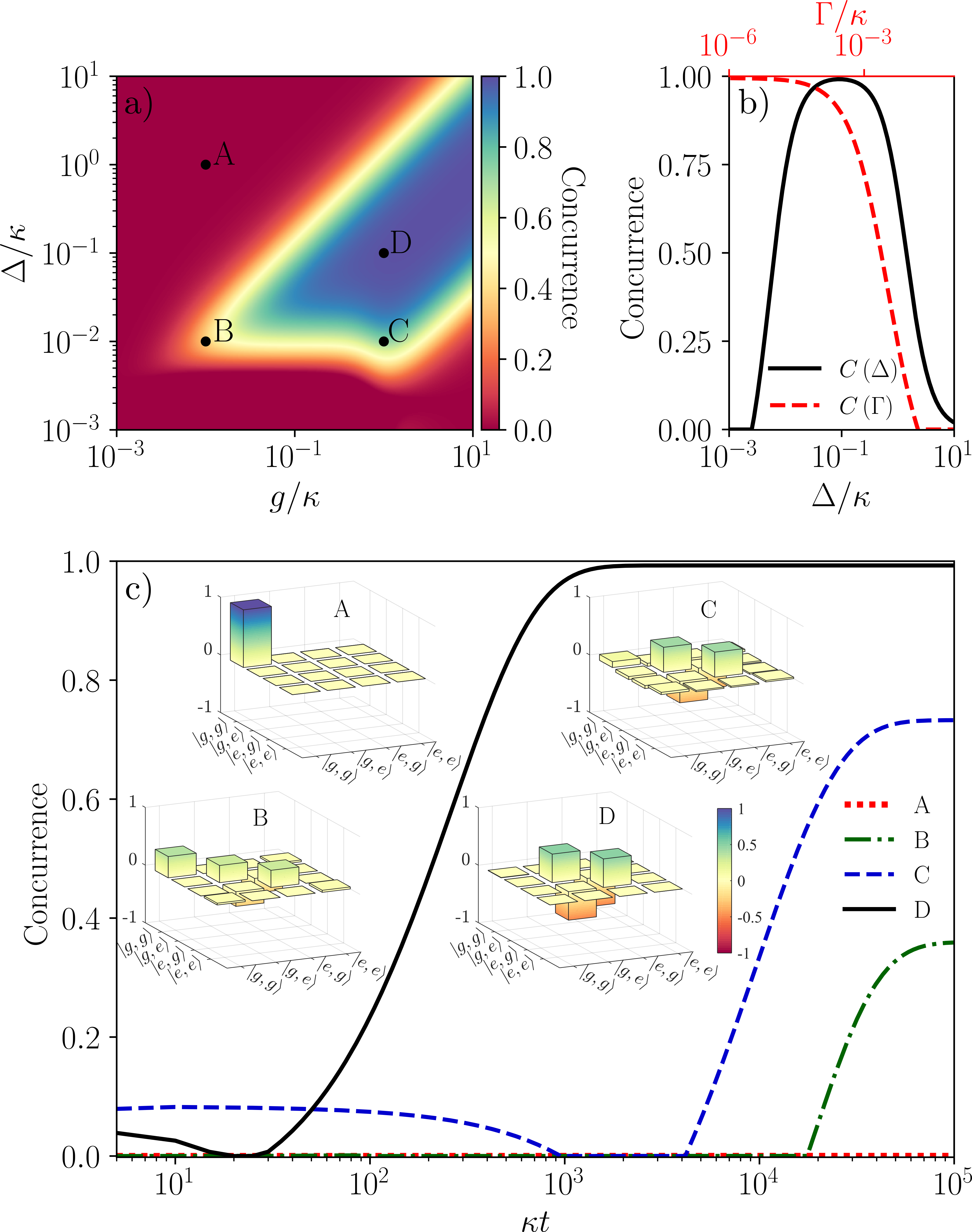}
\par\end{centering}
\caption{Entanglement generation using the steady-state method. (a) Colormap showing the concurrence of the steady state as a function of $\Delta/\kappa$ and $g/\kappa$. Points A to D are referred to in panel (c). (b) Concurrence as a function of the spontaneous decay rate $\Gamma$ of each of the two qubits (top x-axis) and as a function of the detuning (bottom x-axis). (c) Concurrence as a function of time for different parameter sets, given in panel (a). The concurrence generally grows and then stabilises. The insets in panel (c) show the steady state partial density matrix of each curve, where the mode has been traced out. In all panels, the pump strength was set to $\varepsilon=\kappa$, and for (a) and (c) the spontaneous decay rate of each qubit was $\Gamma=10^{-5}\kappa$.}
\label{fig:2}
\end{figure}

As shown in Fig.~\ref{fig:2}(a), we obtain a strong entanglement within a large region of parameters $\Delta$ and $g$. Moreover, we see that satisfying the EIT condition $g\gg \Delta$ is a necessary but not sufficient condition to reach our maximally entangled target state. We have chosen four sets of parameters to illustrate the system dynamics: $A$ shows the situation where $g \ll \Delta$ and the population is lead to the respective dark state, which has a $\vert G\rangle$ character [see Eq.~\eqref{eq:darkstate}], and no entanglement is observed; in $B$, we have $g = \Delta$ which leads to weakly entangled dark state given by a mixture of $\vert g,g\rangle$ and $\vert \Phi_- \rangle$; in $C$ we have an interesting situation where the EIT condition $g \gg \Delta$ is fulfilled, the dark state a $\vert\Phi_- \rangle$ character, but the detuning is so small that it takes too long to populate $\vert \Phi_- \rangle$, and since the system keeps decaying to the true ground state $\vert G \rangle \otimes \vert 0\rangle$, the entanglement is affected; $D$ shows a near-optimal situation, with $g \gg \Delta$ and $\Delta \gg \Gamma$, driving approximately all the population to the dark state $\vert \Phi_- \rangle$.     

In Fig.~\ref{fig:2}(b) we show the behavior of the concurrence as a function of the detuning for a fixed value of $g=\kappa$. We observe that the concurrence decreases significantly when we move away from the optimal point $\Delta\approx0.1g$. This finding is in accordance with what was previously discussed for the parameters sets $A$ and $C$. This optimal ratio between $g$ and $\Delta$ is influenced by the system's decay rate $\Gamma$. A reduced value of $\Gamma$ allows us to achieve entanglement for smaller detunings. For $\Gamma=0$, even an infinitesimal detuning would eventually take the system to $\vert \Phi_- \rangle$. Fig.~\ref{fig:2}(b) shows that the spontaneous decay reduce the entanglement even for the optimal case $D$; the concurrence decreases exponentially for $\Gamma>5\times10^{-4}\kappa$.  

Focusing on the entanglement dynamics, we show in Fig.~\ref{fig:2}(c) the concurrence over time for the cases $A$, $B$, $C$ and $D$, where we observe different time scales to achieve the maximum entanglement for each curve, respectively. We also include a visual representation of the time evolved partial density matrix, where the bosonic mode has been traced out, showing that we achieve the maximally entangled target state $\vert\Phi_-\rangle$ for the near-optimal set of parameters $D$. 

\section{Adiabatic process}\label{sec:stirap}

\begin{figure}[htb]
\begin{centering}
\includegraphics[width=1\columnwidth]{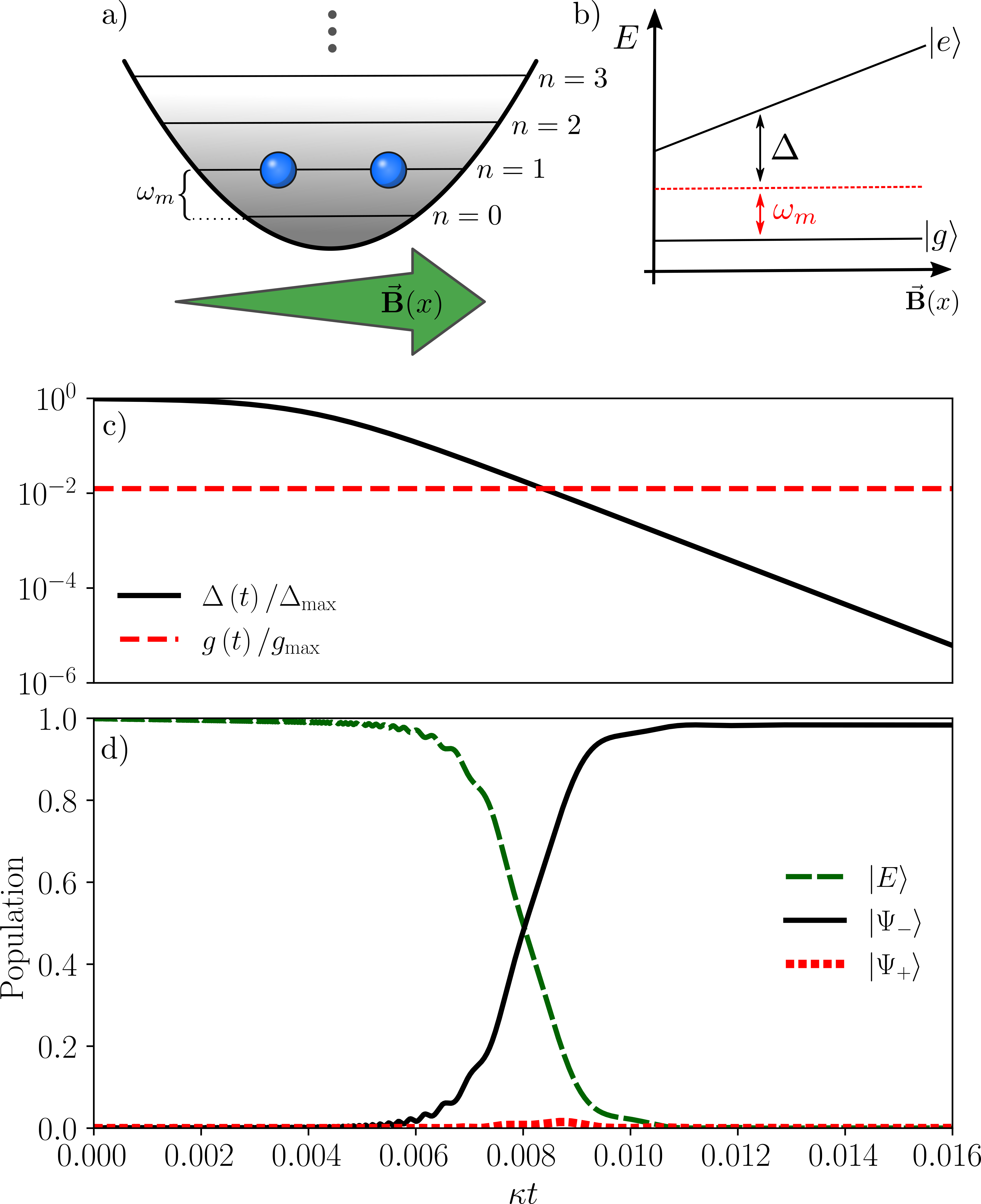}
\par\end{centering}
\caption{Entanglement generation - adiabatic method. (a) Two two-level ions are trapped in a harmonic potential and coupled to the same phonon mode. The ions are subjected to a magnetic field gradient, which promotes different energy shifts to their excited states, recovering the system illustrated in Fig.~\ref{fig:1}(a). (b) Pictorial representation of the energy shifts in the ions excited states due to a magnetic field gradient, as a function of their position in the trap. (c) Time evolution of $g$ and $\Delta$, with $\Delta$ varying accordingly to Eq.~\eqref{eq:D} and a constant $g$. (d) Population changes over time. We considered $\Gamma=10^{-3}\kappa$, $\Delta_{\mathrm{max}}=2\times10^{5}\kappa$, $g=2.5\times10^{2}\kappa$, $t_0=4\times10^{-3}/\kappa$, $\lambda=5\times10^2\kappa$.}
\label{fig:3alt}
\end{figure}

Inspired by the schemes to counteract decoherence developed for multilevel atoms, we propose a STIRAP-like process to efficiently populate the entangled state $\vert\Psi_{-}\rangle$. To this end, we consider a trapped ion implementation, where two ions are confined in a harmonic potential and coupled to the same motional mode, as depicted in Fig.~\ref{fig:3alt}(a). The ions are subjected to a magnetic field gradient, thus will experience different energy shifts of their excited states, resembling the system illustrated in Fig.~\ref{fig:1}(a). To perform the adiabatic population transfer, the system must be initially prepared in the state $\vert E\rangle$, which consists of the two ions in the ground state and the motional state with one excitation; the latter can be prepared by exciting one of the ions and then by letting it exchange energy with the vibration mode via a red sideband interaction. The initial state $\vert E\rangle$ corresponds to the dark state when $\Delta\gg g$. By reversing this condition to $g\gg\Delta$ the dark state adiabatically transforms to $\vert\Psi_-\rangle$. By manipulating the coupling strength $g$ and the detuning $\Delta$, which can be accomplished by changing the power of a laser resonant to the red sideband transition and by magnetic field gradient varying in time, respectively, we can control the adiabaticity throughout the process so that all the population is transferred coherently to the final maximally entangled state $\vert\Psi_{-}\rangle$, with a negligible population in the state $\vert\Psi_{+}\rangle$ (see Appendix A). 

This scheme can also be applied to other experimental platforms. However, these might not allow a perfect control over both parameters at the same time, e.g., a fixed $g$ and a tunable $\Delta$, as often observed in quantum circuits. To show that even in this situation we can achieve a high degree of entanglement, we assume $g$ to be constant and we vary $\Delta$ in time, according to 
\begin{equation}
\Delta\left(t\right) = \frac{\Delta_{\mathrm{max}}}{2}\left[1-\tanh{\lambda(t-t_0)}\right], \label{eq:D}    
\end{equation}
as shown in Fig.~\ref{fig:3alt}(c). Here, $\Delta_{\mathrm{max}}$ corresponds to the maximum value of the detuning and $t_0$ is the time at which this function reaches its respective half maximum value $\Delta_{\mathrm{max}}/2$. The adiabaticity of the process is controlled by $\lambda$, which determines the time scale of the parameter swap and, consequently, how fast the parameters are changed. The specific choice of function is not crucial to the method, as long as it guarantees an near-adiabatic inversion of parameters. Setting the value of $g$ that ensures the initial and final conditions $\Delta \gg g$ and $g \gg \Delta$, respectively, we make the parameter swapping as smooth as possible. On the other hand, the lack of control over $g$ restricts the time scale of the population transfer to maintain the adiabaticity during the parameter swap. Nevertheless, we show that one can still perform a STIRAP-like process, obtaining a highly entangled final state, as shown in \ref{fig:3alt}(d), with a negligible population of the state $\vert \Psi_+\rangle$. In the case where both $g$ and $\Delta$ can be simultaneously controlled, a near-perfect STIRAP process can be achieved (see Appendix A).
 
\section{Conclusion}

In conclusion, here we present two novel strategies for producing maximally entangled states in a system of two non-degenerate qubits coupled to a single bosonic mode. In both cases, we use a direct analogy with ordinary three-level atomic systems in the $\Lambda$-configuration and the EIT phenomenon, which allows us to draw a parallel with the processes of optical pumping and adiabatic population transfer.

In the first proposed scheme, we show, that it is possible to generate steady state entanglement with concurrence $C>0.99$. Moreover, we show that the symmetry between the qubits with respect to the bosonic mode is beneficial for the generation of entanglement. As for the second scheme, we generate a highly entangled state by means of an adiabatic process, and we achieve a population over $99\%$ in the state $\vert \Psi_-\rangle$. We emphasize that this was done even considering non-ideal situations, where there is no complete control over $g$ and $\Delta$. In ideal cases, where both parameters can be controlled simultaneously, the adiabatic process can be controlled more efficiently, which leads to a perfectly coherent transfer of populations, as shown in detail in Appendix A.

The results presented in this work, besides indicating new ways to generate highly entangled states in a simple way, have potential application in several experimental platforms, such as trapped ions and quantum dots molecules coupled to a cavity mode.

\acknowledgments
C.J.V.-B., R.B. and M.H.O. thank the support from the National Council for Scientific and Technological Development (CNPq) grants 311612/2021-0, 141247/2018-5, 409946/2018-4 and 313886/2020-2, and from the S\~ao Paulo Research Foundation (FAPESP) through Grants No.~2020/00725-9, 2019/13143-0, 2019/11999-5 and 2018/15554-5. The authors also thank the Joint Brazilian-Swedish Research Collaboration (CAPES-STINT), grant 88887.304806/2018-00 and BR2018-8054. M.H., G.H., C.Z. thank the support by the Knut \& Alice Wallenberg Foundation (through the Wallenberg Centre for Quantum Technology [WACQT]), by Olle Enkvists Stiftelse (Project No 204-0283), by Carl Trygger Stiftelse Grant No. CTS 19-139, and by the Swedish Research Council (Trapped Rydberg Ion Quantum Simulator).

\bibliographystyle{unsrt}
\bibliography{ref}

\begin{thebibliography}{10}

\bibitem{horodecki2009}
Ryszard Horodecki, Pawe{\l} Horodecki, Micha{\l} Horodecki, and Karol
  Horodecki.
\newblock Quantum entanglement.
\newblock {\em Reviews of modern physics}, 81(2):865, 2009.

\bibitem{ekert1991}
Artur~K Ekert.
\newblock Quantum cryptography based on bell’s theorem.
\newblock {\em Physical review letters}, 67(6):661, 1991.

\bibitem{bennett1993}
Charles~H Bennett, Gilles Brassard, Claude Cr{\'e}peau, Richard Jozsa, Asher
  Peres, and William~K Wootters.
\newblock Teleporting an unknown quantum state via dual classical and
  einstein-podolsky-rosen channels.
\newblock {\em Physical review letters}, 70(13):1895, 1993.

\bibitem{bennett2000}
Charles~H Bennett and David~P DiVincenzo.
\newblock Quantum information and computation.
\newblock {\em nature}, 404(6775):247--255, 2000.

\bibitem{gottesman1999}
Daniel Gottesman and Isaac~L Chuang.
\newblock Demonstrating the viability of universal quantum computation using
  teleportation and single-qubit operations.
\newblock {\em Nature}, 402(6760):390--393, 1999.

\bibitem{giovannetti2006}
Vittorio Giovannetti, Seth Lloyd, and Lorenzo Maccone.
\newblock Quantum metrology.
\newblock {\em Physical review letters}, 96(1):010401, 2006.

\bibitem{degen2017}
Christian~L Degen, F~Reinhard, and Paola Cappellaro.
\newblock Quantum sensing.
\newblock {\em Reviews of modern physics}, 89(3):035002, 2017.

\bibitem{lidar1998}
Daniel~A Lidar, Isaac~L Chuang, and K~Birgitta Whaley.
\newblock Decoherence-free subspaces for quantum computation.
\newblock {\em Physical Review Letters}, 81(12):2594, 1998.

\bibitem{viola1998}
Lorenza Viola and Seth Lloyd.
\newblock Dynamical suppression of decoherence in two-state quantum systems.
\newblock {\em Phys. Rev. A}, 58:2733--2744, Oct 1998.

\bibitem{steane1996}
Andrew~M Steane.
\newblock Error correcting codes in quantum theory.
\newblock {\em Physical Review Letters}, 77(5):793, 1996.

\bibitem{knill1997}
Emanuel Knill and Raymond Laflamme.
\newblock Theory of quantum error-correcting codes.
\newblock {\em Phys. Rev. A}, 55:900--911, Feb 1997.

\bibitem{sun2010}
Qingqing Sun, M.~Al-Amri, Luiz Davidovich, and M.~Suhail Zubairy.
\newblock Reversing entanglement change by a weak measurement.
\newblock {\em Phys. Rev. A}, 82:052323, Nov 2010.

\bibitem{kim2012}
Yong-Su Kim, Jong-Chan Lee, Osung Kwon, and Yoon-Ho Kim.
\newblock Protecting entanglement from decoherence using weak measurement and
  quantum measurement reversal.
\newblock {\em Nature Physics}, 8(2):117--120, 2012.

\bibitem{facchi2004}
Paolo Facchi, DA~Lidar, and S~Pascazio.
\newblock Unification of dynamical decoupling and the quantum zeno effect.
\newblock {\em Physical Review A}, 69(3):032314, 2004.

\bibitem{maniscalco2008}
Sabrina Maniscalco, Francesco Francica, Rosa~L Zaffino, Nicola~Lo Gullo, and
  Francesco Plastina.
\newblock Protecting entanglement via the quantum zeno effect.
\newblock {\em Physical review letters}, 100(9):090503, 2008.

\bibitem{diehl2008}
Sebastian Diehl, A~Micheli, Adrian Kantian, B~Kraus, HP~B{\"u}chler, and Peter
  Zoller.
\newblock Quantum states and phases in driven open quantum systems with cold
  atoms.
\newblock {\em Nature Physics}, 4(11):878--883, 2008.

\bibitem{kraus2008}
Barbara Kraus, Hans~P B{\"u}chler, Sebastian Diehl, Adrian Kantian, Andrea
  Micheli, and Peter Zoller.
\newblock Preparation of entangled states by quantum markov processes.
\newblock {\em Physical Review A}, 78(4):042307, 2008.

\bibitem{verstraete2009}
Frank Verstraete, Michael~M Wolf, and J~Ignacio Cirac.
\newblock Quantum computation and quantum-state engineering driven by
  dissipation.
\newblock {\em Nature physics}, 5(9):633--636, 2009.

\bibitem{plenio1999}
Martin~B Plenio, SF~Huelga, A~Beige, and PL~Knight.
\newblock Cavity-loss-induced generation of entangled atoms.
\newblock {\em Physical Review A}, 59(3):2468, 1999.

\bibitem{nicolosi2004}
S~Nicolosi, A~Napoli, A~Messina, and F~Petruccione.
\newblock Dissipation-induced stationary entanglement in dipole-dipole
  interacting atomic samples.
\newblock {\em Physical Review A}, 70(2):022511, 2004.

\bibitem{wang2010}
Xiaoting Wang and Sophie~G Schirmer.
\newblock Generating maximal entanglement between non-interacting atoms by
  collective decay and symmetry breaking.
\newblock {\em arXiv preprint arXiv:1005.2114}, 2010.

\bibitem{kastoryano2011}
Michael~James Kastoryano, Florentin Reiter, and Anders~S{\o}ndberg S{\o}rensen.
\newblock Dissipative preparation of entanglement in optical cavities.
\newblock {\em Physical review letters}, 106(9):090502, 2011.

\bibitem{del2011}
Elena Del~Valle.
\newblock Steady-state entanglement of two coupled qubits.
\newblock {\em JOSA B}, 28(2):228--235, 2011.

\bibitem{shankar2013}
Shyam Shankar, Michael Hatridge, Zaki Leghtas, KM~Sliwa, Aniruth Narla, Uri
  Vool, Steven~M Girvin, Luigi Frunzio, Mazyar Mirrahimi, and Michel~H Devoret.
\newblock Autonomously stabilized entanglement between two superconducting
  quantum bits.
\newblock {\em Nature}, 504(7480):419--422, 2013.

\bibitem{kimchi2016}
ME~Kimchi-Schwartz, L~Martin, E~Flurin, C~Aron, M~Kulkarni, HE~Tureci, and
  I~Siddiqi.
\newblock Stabilizing entanglement via symmetry-selective bath engineering in
  superconducting qubits.
\newblock {\em Physical review letters}, 116(24):240503, 2016.

\bibitem{liu2016}
Yehan Liu, Shyam Shankar, Nissim Ofek, Michael Hatridge, Anirudh Narla,
  KM~Sliwa, Luigi Frunzio, Robert~J Schoelkopf, and Michel~H Devoret.
\newblock Comparing and combining measurement-based and driven-dissipative
  entanglement stabilization.
\newblock {\em Physical Review X}, 6(1):011022, 2016.

\bibitem{doucet2020}
E~Doucet, F~Reiter, L~Ranzani, and A~Kamal.
\newblock High fidelity dissipation engineering using parametric interactions.
\newblock {\em Physical Review Research}, 2(2):023370, 2020.

\bibitem{krauter2011}
Hanna Krauter, Christine~A Muschik, Kasper Jensen, Wojciech Wasilewski, Jonas~M
  Petersen, J~Ignacio Cirac, and Eugene~S Polzik.
\newblock Entanglement generated by dissipation and steady state entanglement
  of two macroscopic objects.
\newblock {\em Physical review letters}, 107(8):080503, 2011.

\bibitem{santos2022}
Alan~C Santos, Andr{\'e} Cidrim, Celso~J Villas-Boas, Robin Kaiser, and Romain
  Bachelard.
\newblock Generating long-lived entangled states with free-space collective
  spontaneous emission.
\newblock {\em arXiv preprint arXiv:2110.15033}, 2021.

\bibitem{rao2013}
DD~Bhaktavatsala Rao and Klaus M{\o}lmer.
\newblock Dark entangled steady states of interacting rydberg atoms.
\newblock {\em Physical review letters}, 111(3):033606, 2013.

\bibitem{carr2013}
A.~W. Carr and M.~Saffman.
\newblock Preparation of entangled and antiferromagnetic states by dissipative
  rydberg pumping.
\newblock {\em Phys. Rev. Lett.}, 111:033607, Jul 2013.

\bibitem{shao2017}
XQ~Shao, JH~Wu, XX~Yi, and Gui-Lu Long.
\newblock Dissipative preparation of steady greenberger-horne-zeilinger states
  for rydberg atoms with quantum zeno dynamics.
\newblock {\em Physical Review A}, 96(6):062315, 2017.

\bibitem{barreiro2011}
Julio~T Barreiro, Markus M{\"u}ller, Philipp Schindler, Daniel Nigg, Thomas
  Monz, Michael Chwalla, Markus Hennrich, Christian~F Roos, Peter Zoller, and
  Rainer Blatt.
\newblock An open-system quantum simulator with trapped ions.
\newblock {\em Nature}, 470(7335):486--491, 2011.

\bibitem{lin2013}
Yiheng Lin, JP~Gaebler, Florentin Reiter, Ting~Rei Tan, Ryan Bowler,
  AS~S{\o}rensen, Dietrich Leibfried, and David~J Wineland.
\newblock Dissipative production of a maximally entangled steady state of two
  quantum bits.
\newblock {\em Nature}, 504(7480):415--418, 2013.

\bibitem{bentley2014}
CDB Bentley, ARR Carvalho, D~Kielpinski, and JJ~Hope.
\newblock Detection-enhanced steady state entanglement with ions.
\newblock {\em Physical Review Letters}, 113(4):040501, 2014.

\bibitem{horn2018}
Karl~P Horn, Florentin Reiter, Yiheng Lin, Dietrich Leibfried, and Christiane~P
  Koch.
\newblock Quantum optimal control of the dissipative production of a maximally
  entangled state.
\newblock {\em New Journal of Physics}, 20(12):123010, 2018.

\bibitem{cole2021}
Daniel~C Cole, Jenny~J Wu, Stephen~D Erickson, Pan-Yu Hou, Andrew~C Wilson,
  Dietrich Leibfried, and Florentin Reiter.
\newblock Dissipative preparation of w states in trapped ion systems.
\newblock {\em New Journal of Physics}, 23(7):073001, 2021.

\bibitem{cole2022}
Daniel~C Cole, Stephen~D Erickson, Giorgio Zarantonello, Karl~P Horn, Pan-Yu
  Hou, Jenny~J Wu, Daniel~H Slichter, Florentin Reiter, Christiane~P Koch, and
  Dietrich Leibfried.
\newblock Resource-efficient dissipative entanglement of two trapped-ion
  qubits.
\newblock {\em Physical Review Letters}, 128(8):080502, 2022.

\bibitem{vitanov2017}
Nikolay~V. Vitanov, Andon~A. Rangelov, Bruce~W. Shore, and Klaas Bergmann.
\newblock Stimulated raman adiabatic passage in physics, chemistry, and beyond.
\newblock {\em Rev. Mod. Phys.}, 89:015006, Mar 2017.

\bibitem{bergmann2019}
Klaas Bergmann, Hanns-Christoph Nägerl, Cristian Panda, Gerald Gabrielse,
  Eduard Miloglyadov, Martin Quack, Georg Seyfang, Gunther Wichmann, Silke
  Ospelkaus, Axel Kuhn, Stefano Longhi, Alexander Szameit, Philipp Pirro,
  Burkard Hillebrands, Xue-Feng Zhu, Jie Zhu, Michael Drewsen, Winfried~K
  Hensinger, Sebastian Weidt, Thomas Halfmann, Hai-Lin Wang, Gheorghe~Sorin
  Paraoanu, Nikolay~V Vitanov, Jordi Mompart, Thomas Busch, Timothy~J Barnum,
  David~D Grimes, Robert~W Field, Mark~G Raizen, Edvardas Narevicius, Marcis
  Auzinsh, Dmitry Budker, Adriana P{\'{a}}lffy, and Christoph~H Keitel.
\newblock Roadmap on {STIRAP} applications.
\newblock {\em Journal of Physics B: Atomic, Molecular and Optical Physics},
  52(20):202001, sep 2019.

\bibitem{hettich2002}
C.~Hettich, C.~Schmitt, J.~Zitzmann, S.~Kühn, I.~Gerhardt, and V.~Sandoghdar.
\newblock Nanometer resolution and coherent optical dipole coupling of two
  individual molecules.
\newblock {\em Science}, 298(5592):385--389, 2002.

\bibitem{trebbia2021}
J-B Trebbia, Q~Deplano, P~Tamarat, and B~Lounis.
\newblock Tailoring the degree of entanglement of two coherently coupled
  quantum emitters.
\newblock {\em arXiv preprint arXiv:2109.10584}, 2021.

\bibitem{fleischhauer2005}
Michael Fleischhauer, Atac Imamoglu, and Jonathan~P Marangos.
\newblock Electromagnetically induced transparency: Optics in coherent media.
\newblock {\em Reviews of modern physics}, 77(2):633, 2005.

\bibitem{petruccione2002}
H.P. Breuer, F.~Petruccione, and S.P.A.P.F. Petruccione.
\newblock {\em The Theory of Open Quantum Systems}.
\newblock Oxford University Press, 2002.

\bibitem{Wootters1997}
Scott Hill and William~K. Wootters.
\newblock Entanglement of a pair of quantum bits.
\newblock {\em Phys. Rev. Lett.}, 78:5022--5025, Jun 1997.

\bibitem{Wootters1998}
William~K. Wootters.
\newblock Entanglement of formation of an arbitrary state of two qubits.
\newblock {\em Phys. Rev. Lett.}, 80:2245--2248, Mar 1998.

\bibitem{qutip2013}
J.R. Johansson, P.D. Nation, and Franco Nori.
\newblock Qutip 2: A python framework for the dynamics of open quantum systems.
\newblock {\em Computer Physics Communications}, 184(4):1234--1240, 2013.

\end{thebibliography}

\appendix

\section{Perfect parameter controllability in the adiabatic scheme}

As shown in Sec.~\ref{sec:stirap}, the proposed adiabatic scheme does not require a perfect simultaneous control over $g$ and $\Delta$ in order to achieve a highly entangled steady state. However, if the system allows this level of controllability, it is possible to guarantee that the inversion of parameters occurs almost in a perfect adiabatic way and on shorter time scales. In this case, practically all the population is transferred coherently in the system from the initial dark state $\vert E\rangle$ to the final one $\vert\Psi_{-}\rangle$, with a negligible population in the excited state. 

Let us consider the case of trapped ions, as described in Sec.~\ref{sec:stirap} and depicted in Fig.~\ref{fig:3alt}(a). In this scheme, we consider that both $\Delta$ and $g$ can be externally manipulated at the same time. To illustrate, we here consider the time variation of $\Delta$ to be still dictated by Eq.~\eqref{eq:D} and $g$ to vary as
\begin{equation}
g\left(t\right)  =  \frac{g_\mathrm{max}}{2}\left[1+\tanh{\lambda(t-t_0)}\right],\label{caseg}\\
\end{equation}
where $g_{\mathrm{max}}$ is the maximum value of the coupling constant, and $t_0$ and $\lambda$ are the same parameters from Eq.~\eqref{eq:D}.

\begin{figure}[htbp]
\begin{centering}
\includegraphics[width=1\columnwidth]{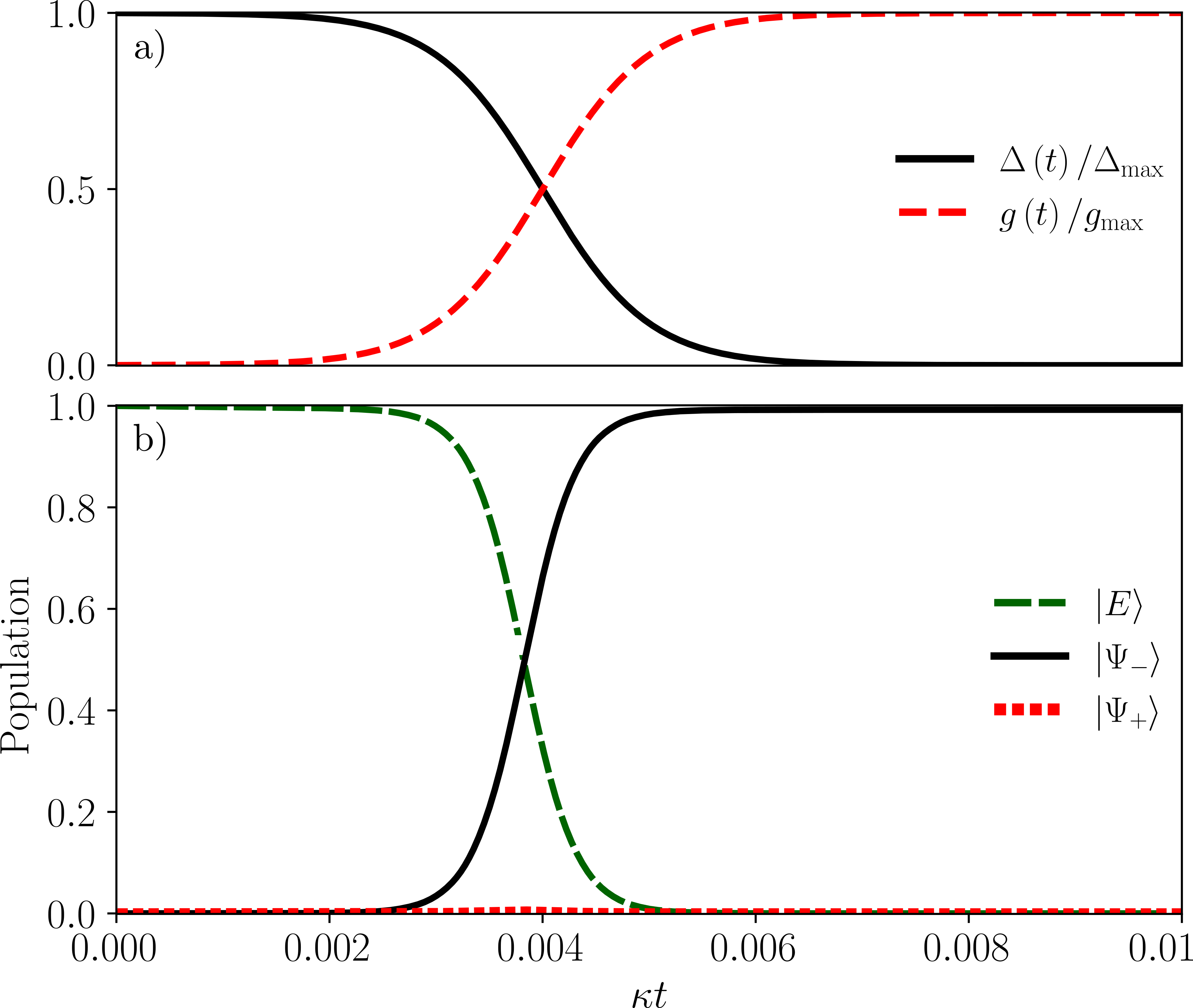}
\par\end{centering}
\caption{Entanglement generation - adiabatic method with tunable $g$ and $\Delta$. (a) Time evolution of the parameters $g\left(t\right)$ and $\Delta\left(t\right)$ as described in Eq.~\eqref{caseg} and \eqref{eq:D}, with the corresponding populations changes in (b). We considered  for all simulations. Adopted parameters: $\Gamma=10^{-3}\kappa$, $g_{\mathrm{max}}=\Delta_{\mathrm{max}}=2\times10^{4}\kappa$, $t_0=3\times10^{-3}\kappa$, $\lambda=10^3\kappa$. }
\label{fig:3}
\end{figure}

In Fig.~\ref{fig:3}(b), we can see how the populations in the three states of the single excitation subspace evolve through time, showing that we coherently transfer the population from $\vert E \rangle$ to $\vert \Psi_-\rangle$ ($\langle \vert \Psi_-\rangle\langle\Psi_-\vert\rangle>0.99$) before the mode dissipation starts to be relevant. The parameters $\Delta$ and $g$ were changed according to Eqs. \eqref{eq:D} and \eqref{caseg} , as shown in Fig.~\ref{fig:3}(a).

\end{document}